\documentclass[aps,physrev,twocolumn,groupedaddress]{revtex4-2}
\pdfoutput=1
\usepackage[utf8]{inputenc}
\usepackage[english]{babel}
\usepackage[T1]{fontenc}
\usepackage{amsmath}
\usepackage{amssymb}
\usepackage{amsthm}
\usepackage{hyperref}
\usepackage{braket} 

\usepackage{tikz}
\usepackage{lipsum}

\newtheorem{theorem}{Theorem}
\theoremstyle{definition}
\newtheorem{definition}{Definition}

\begin{document}

\title{Probability-Phase Mutual Information}

\author{Cameron Hahn}

\author{Nishan Ranabhat}

\author{Fabio Anza}

\affiliation{Department of Physics, University of Maryland, Baltimore County, Baltimore, MD 21250, USA}
\affiliation{Quantum Science Institute, University of Maryland, Baltimore County, Baltimore, MD 21250, USA}


\begin{abstract}
Quantum coherence is an exquisitely quantum phenomenon that depends on both probability amplitudes and relative phases. Standard coherence measures quantify superposition within density matrices but cannot distinguish ensembles that produce the same mixed state through different distributions of pure states. Building on the geometric formulation of quantum mechanics, we introduce the probability-phase mutual information $I(P;\Phi)$. We show that it characterizes quantum coherence at the ensemble level and that ensemble coherence systematically exceeds density-matrix coherence, thus quantifying the structure lost when averaging over pure states. Eventually, its relevance for quantum thermodynamics, quantum information theory, and deep thermalization is highlighted by explicit examples: canonical ensembles reveal temperature-dependent probability-phase correlations absent from thermal density matrices; we show that the probability of converting an ensemble into another one is bound by the ratio of their $I(P;\Phi)$; and, that a non-vanishing $I(P;\Phi)$ signals the breakdown of deep thermalization.
\end{abstract}

\maketitle

\section{Introduction}
\label{sec:introduction}

Quantum coherence is one of the central resources that distinguishes quantum mechanics from classical physics. 
It underlies the advantage of many quantum technologies, ranging from computation and metrology to quantum state interconversion, and has been formalized through the 
resource theory of coherence~\cite{Baumgratz2014QuantifyingCoherence,Streltsov2017QuantumCoherenceResource}. In this standard framework, coherence is 
quantified at the level of density matrices, with measures such as the relative entropy of coherence $\mathcal{C}(\rho)$ \cite{herbut_quantum_2005} capturing the degree to which a quantum state is a superposition in a chosen basis. 

While powerful, this perspective overlooks a different layer of structure: the correlations 
that can exist within a  collection of pure states––an ensemble––whose average is the density matrix. While two ensembles may have the same density matrix, they can differ sharply in how their pure states are distributed. Standard coherence measures can not distinguish such cases, leaving ensemble-level coherence phenomena outside their scope. Ensemble-level correlations have recently garnered significant attention through the study of deep thermalization \cite{Ippoliti2022SolvableModelDeep, Mark2024MaximumEntropyDeep, Rigol2010QuantumChaosThermalization, Lucas2023GeneralizedDeepThermalization, Ippoliti2023DynamicalPurificationEmergence, Bhore2023DeepThermalizationConstrained, Ho2022ExactEmergentQuantum}, which has recently been physically realized in neutral atom quantum simulators \cite{choiPreparingRandomStates2023, cotlerEmergentQuantumState2023}. In this phenomenon, the projected ensemble of pure states obtained by measuring a subsystem's complement converges (in chaotic systems) to the Haar-uniform distribution, a statement that goes strictly beyond the relaxation of the density matrix to a thermal state, and is not fully captured by any finite-order moment of the density matrix. This motivates treating the ensemble itself, rather than the density matrix it generates, as the primary object of study.


Geometric quantum mechanics (GQM) \cite{Strocchi1966ComplexCoordinatesQuantum,Mielnik1968GeometryQuantumStates,Kibble1979GeometrizationQuantumMechanics,Heslot1985QuantumMechanicsClassical,Page1987GeometricalBerryPhase,Anandan1990GeometryQuantumEvolution,Gibbons1992TypicalStatesDensity,Ashtekar1995GeometryQuantumMechanics,Ashtekar1999GeometricalFormulationQuantum,Brody2001GeometricQuantumMechanics,Carinena2007GeometrizationQuantumMechanics,Chruscinski2006GeometricAspectsQuantum,Marmo2010GeometricalDescriptionQuantum,Avron2019ElementaryIntroductionGeometry,Pastorello2015GeometricHamiltonianDescription,Pastorello2015GeometricHamiltonianFormulation,Pastorello2016GeometricHamiltonianApplications,Clemente2013EhrenfestPictureGeometry} provides a natural setting in which to formulate a theory of coherence for ensembles, as it enables us to consider the full distribution rather than moments of the density matrix. In a chosen basis, the complex projective space of pure state $\mathbb{C}P^{D-1}$ is coordinatized by probabilities $p_k$ (encoding measurement-accessible statistics in the chosen basis) 
and phases $\phi_k$ (carrying the complementary, measurement-inaccessible information). This accessible–inaccessible dichotomy suggests a fundamental question: how much information about the phases is statistically tied to the probabilities 
across an ensemble?

In this work we answer this question by introducing the \emph{probability–phase mutual information} $I(P;\Phi)$ and showing that it is a measure of coherence for ensembles of pure states. Using a resource-theoretic approach, our first result establishes that $I(P;\Phi)$ satisfies all six axioms \cite{Streltsov2017QuantumCoherenceResource} of coherence measures (non-negativity, monotonicity, strong monotonicity, convexity, additivity, and pure-state normalization), thereby qualifying as a coherence measure for ensembles. Our second result establishes the relationship to the standard relative entropy of coherence $\mathcal{C}(\rho)$ through the \emph{coherence surplus}, revealing how ensembles, and their coherence, relate to density-matrix coherence. Our final results make the operational meaning of $I(P;\Phi)$ clear: it bounds the probability to convert one ensemble into another via stochastic free operations.

The paper is organized as follows. In Section \ref{sec:prob-phase-mutual-info} we provide a brief summary of geometric quantum mechanics and establish the mathematical aspects of probability-phase mutual information. In Section \ref{sec:prob-phase-mutual-info-as-coherence} we develop the resource theory of coherence for ensembles, defining incoherent states and free operations, and proving that $I(P;\Phi)$ satisfies all standard coherence axioms. In Section \ref{sec:examples} we present analytical and numerical examples demonstrating the behavior of $I(P;\Phi)$ across diverse geometric quantum states and physical situations. In Section \ref{sec:discussion} we discuss the relationship between $I(P;\Phi)$ and the standard relative entropy of coherence $\mathcal{C}(\rho)$. We then introduce the coherence surplus and highlight their complementary operational meanings, establishing a state conversion bound. Finally, Section~\ref{sec:conclusions} concludes with a summary of results and future directions. Detailed proofs of the coherence axioms and the non-negativity of the coherence surplus are provided in Appendix~\ref{appendix} and \ref{apx:coherence-theorem}, respectively.

\section{Probability-Phase Mutual Information}
\label{sec:prob-phase-mutual-info}

Geometric quantum mechanics (GQM) \cite{Strocchi1966ComplexCoordinatesQuantum,Mielnik1968GeometryQuantumStates,Kibble1979GeometrizationQuantumMechanics,Heslot1985QuantumMechanicsClassical,Page1987GeometricalBerryPhase,Anandan1990GeometryQuantumEvolution,Gibbons1992TypicalStatesDensity,Ashtekar1995GeometryQuantumMechanics,Ashtekar1999GeometricalFormulationQuantum,Brody2001GeometricQuantumMechanics,Carinena2007GeometrizationQuantumMechanics,Chruscinski2006GeometricAspectsQuantum,Marmo2010GeometricalDescriptionQuantum,Avron2019ElementaryIntroductionGeometry,Pastorello2015GeometricHamiltonianDescription,Pastorello2015GeometricHamiltonianFormulation,Pastorello2016GeometricHamiltonianApplications,Clemente2013EhrenfestPictureGeometry} provides a differential-geometric framework for quantum systems. For a quantum system with Hilbert space $\mathcal{H}$ of dimension $D$, the space of pure states is the complex projective space $\mathcal{P}(\mathcal{H}) \sim \mathbb{C}P^{D-1}$. This identification removes the redundancy inherent in the standard formulation---vectors differing only by normalization and global phase represent the same physical state.

The complex projective space $\mathbb{C}P^{D-1}$ is a Kähler manifold, equipped with both a Riemannian geometry–providing the Fubini-Study metric–and a symplectic geometry–providing a closed non-degenerate symplectic two-form\cite{Bengtsson2006GeometryQuantumStates,chern_complex_1979}. The existence of a symplectic structure is the hallmark of Hamiltonian mechanics, and it underlies a deep geometric parallel between classical and quantum mechanics that, while known \cite{Strocchi1966ComplexCoordinatesQuantum,Kibble1979GeometrizationQuantumMechanics,Ashtekar1995GeometryQuantumMechanics,Heslot1985QuantumMechanicsClassical} goes often underappreciated.

In this setting, probability and phase coordinates arise naturally after a basis' choice $\{\ket{e_k}\}_{k=0}^{D-1}$ in the Hilbert space $\mathcal{H}$. They provide a natural parametrization of $\mathbb{C}P^{D-1}$ through $Z^k = \sqrt{p_k} e^{i\phi_k}$, where $\ket{\psi} = \sum_{k=0}^{D-1} Z^k \ket{e_k}$ and we set $\phi_0 = 0$ without loss of generality ~\cite{Anza2021DensityMatricesGeometric}. A point in the manifold of quantum states is therefore identified by $p = (p_1, \ldots, p_{D-1}) \in \Sigma_{D-1}$, where $\Sigma_{D-1}$ is the $(D-1)$-dimensional probability simplex defined by $p_k \geq 0$ and $\sum_{k=0}^{D-1} p_k = 1$ (with $p_0 = 1 - \sum_{k=1}^{D-1} p_k$), and $\phi = (\phi_1, \ldots, \phi_{D-1}) \in \mathbb{T}_{D-1}$, the $(D-1)$-dimensional torus with $\phi_k \in [0, 2\pi)$. This choice diagonalizes the symplectic structure and reveals the canonical nature of $(p,\phi)$ coordinates.

Ensembles are introduced by promoting the pure state of a quantum system to a random variable $\Psi$, distributed according to some probability measure $\mu_\Psi$ on $\mathcal{P}(\mathcal{H})$. We call $\mu_{P,\Phi}$ a \emph{geometric quantum state} \cite{Anza2021DensityMatricesGeometric}. While the measure is invariant under changes of coordinates, here we adopt a coordinate-specific notation to address the statistical correlations between probabilities and phases: we introduce random variables $(P,\Phi)$, with $\Psi(P,\Phi) = \sqrt{P}e^{i\Phi}$ and realizations $(p,\phi) \in \mathbb{C}P^{D-1}$, distributed according to $\mu_{P,\Phi}$. 

When $\mu_{P,\Phi}$ is absolutely continuous with respect to the Fubini-Study volume measure $dV_{FS}$ on $\mathcal{P}(\mathcal{H})$, it can be expressed through a probability density $q(p,\phi)$ such that:
\begin{align}
\mu_{P,\Phi}(A) = \int_A q(p,\phi) \, dV_{FS} \nonumber
\end{align}
for any measurable set $A \subseteq \mathcal{P}(\mathcal{H})$.

In probability-phase coordinates, the Fubini-Study volume element factorizes as
\begin{align}
dV_{FS} = \prod_{k=1}^{D-1} \frac{dp_k \, d\phi_k}{2} \label{eq:fubini-study-element}
\end{align}
This is formally identical to the position/momentum factorization of the phase space volume element $\prod_i dp_i \, dq_i$ in classical statistical mechanics, where $(p,q)$ are canonically conjugate momentum-position pairs (note that the $dp_k$ in Eq. \ref{eq:fubini-study-element} are probabilities, not momenta).

Given a joint measure $\mu_{P,\Phi}$, the marginals can be obtained through partial integration:
\begin{subequations}
\begin{align}
\mu_P(B) &= \int_{B \times \mathbb{T}_{D-1}} \mu_{P,\Phi}(dp,d\phi) \nonumber \\
\mu_\Phi(C) &= \int_{\Sigma_{D-1} \times C} \mu_{P,\Phi}(dp,d\phi) \nonumber
\end{align}
\end{subequations}
where $B \subseteq \Sigma_{D-1}$ and $C \subseteq \mathbb{T}_{D-1}$ are subsets of the probability simplex and the phase hyper-torus. A product measure $\mu_P \cdot \mu_\Phi$ represents the case where probabilities and phases are statistically independent.

To properly define the probability-phase mutual information in a geometric quantum state, we employ the scaling framework originally developed by Renyi \cite{Renyi1959DimensionEntropyProbability} and extended to quantum systems in Ref.~\cite{Anza2022QuantumInformationDimension}. Calling $Z^\epsilon$ a coarse-graining of the state space at scale $\epsilon$, and $H(Z^\epsilon)$ the Shannon entropy of the resulting discrete distribution, one has
\begin{equation}
    H(Z^\epsilon) \sim_{\epsilon \to 0} \mathfrak{D}(-\log \epsilon) + H_G \label{eq:entropy_scaling}
\end{equation}
where $\mathfrak{D}$ is the \emph{quantum information dimension} and $H_G$ is the \emph{geometric entropy}. For measures with support on a finite number of points $\mu = \sum_{k=1}^{N}q_k \delta_{\chi_\alpha}$ we have $\mathfrak{D}=0$ and $H_G = - \sum_{\alpha=1}^N q_\alpha \log q_\alpha$. Measures with full support on $\mathcal{P}(\mathcal{H})$ exhibit $\mathfrak{D} = 2(D-1)$, and the geometric entropy reduces to the differential entropy:
\begin{align}
H_{2(D-1)}[\mu_{P,\Phi}] = -\int_{\mathcal{P}(\mathcal{H})} \!\!\! q(p,\phi) \log q(p,\phi) \, dV_{FS}~.\label{eq:geometric_entropy}
\end{align}

The marginal entropies follow by applying the scaling method defined in Ref.\cite{Anza2022QuantumInformationDimension} to the marginal measures. Defining $Z^\epsilon_p$ and $Z^\epsilon_\phi$ to be coarse grained variables in $\Sigma_{D-1}$ and $\mathbb{T}_{D-1}$ respectively, we have
\begin{align*}
    & H(Z^\epsilon_P) \sim_{\epsilon \to 0} \mathfrak{D}_P(-\log \epsilon) + H_P \\
    &H(Z^\epsilon_{\Phi}) \sim_{\epsilon \to 0} \mathfrak{D}_\Phi(-\log \epsilon) + H_{\Phi}
\end{align*}
This leads to the definition of mutual information dimension and probability-phase mutual information from the resulting scaling:
\begin{subequations}
\begin{align}
I_\epsilon(P;\Phi) & := H(Z^\epsilon_P) + H(Z^\epsilon_{\Phi}) - H(Z^{\epsilon}_{P,\Phi}) \label{eq:mutual_information_discretized} \\
& \sim_{\epsilon \to 0} \mathfrak{D}_I (-\log \epsilon) + I(P;\Phi)~ \label{eq:mutual_information_scaling}
\end{align}
\end{subequations}
where $\mathfrak{D}_I \equiv \mathfrak{D}_P+\mathfrak{D}_\Phi - \mathfrak{D}_{P,\Phi}$ and, similarly, $I(P,\Phi) \equiv H_P + H_{\Phi} - H_{P,\Phi}$.
The \emph{mutual information dimension} $\mathfrak{D}_I$ measures the dimensionality surplus in the support of $\mu_P \mu_{\Phi}$ versus the true joint $\mu_{P,\Phi}$ while $I(P;\Phi)$ is the \emph{probability-phase mutual information}: it measures the statistical dependence between probability and phase variables in the ensemble described by $\mu_{P,\Phi}$. Equivalently, it can be expressed as the relative entropy:
\begin{align}
I(P;\Phi) = D_{\mathrm{KL}}(\mu_{P,\Phi} || \mu_P \cdot \mu_\Phi) \label{eq:mutual_information_relative_entropy}
\end{align}
quantifying how far the joint distribution deviates from statistical independence. Importantly, we emphasize that $I(P;\Phi)$ is explicitly basis dependent, as the decomposition into probability-phase coordinates is dependent on the choice of basis $\{\ket{e_k}\}$. This basis dependence is a standard feature of all coherence measures which persists at the ensemble-level.

This concludes the mathematical exposition of geometric quantum states and their probability-phase mutual information. Before establishing the formal resource theory, we address a key conceptual question in the next subsection.

\subsection{Why probability-phase mutual information?}

We now address a fundamental question: why is $I(P;\Phi)$ the natural quantity for characterizing coherence at the ensemble level?

Quantum coherence fundamentally arises from superposition—the ability of quantum states to exist in combinations of basis states with both definite probability amplitudes and relative phases. In the geometric framework, these two aspects are separated into probability coordinates $P$ (accessible through measurements in the reference basis) and phase coordinates $\Phi$ (inaccessible to such measurements). 

When probabilities and phases are statistically independent—that is, when $\mu_{P,\Phi} = \mu_P \cdot \mu_\Phi$—knowledge of the probabilities yields no information about the phases. This represents the absence of systematic correlations between accessible and inaccessible degrees of freedom. Conversely, when $I(P;\Phi)$ is large, phases and probabilities are highly correlated.

Consider a simple example: an ensemble of qubit states $\ket{\psi} = \sqrt{1-p}\ket{0} + \sqrt{p}e^{i\phi}\ket{1}$. If $\phi$ is chosen uniformly regardless of $p$, then $I(P;\Phi)=0$ and the ensemble exhibits no systematic phase structure. If instead $\phi = 2\pi p + \epsilon$ (with small noise $\epsilon$), then probabilities and phases are linearly correlated, yielding finite $I(P;\Phi)$ and reflecting structured interference patterns across the ensemble.
\begin{table*}
\centering
{\fontsize{12pt}{14pt}\selectfont
\begin{tabular}{| l | l | l |} 
 \hline
 General Name & QM & GQM  \\ [0.5ex] 
 \hline\hline
 State & $\rho$ – Density Matrix & $\mu$ – Geometric Quantum State  \\ 
 \hline
 Operations & $\Lambda$ – CPTP maps & $\mathcal{K}$ – Geometric Channel \\
 \hline
 Dephasing & $\Delta$ – Dephasing & $\Delta_G$ – Geometric Dephasing\\
 \hline
 Free states & $\Delta[\rho] \in \mathcal{I}$ – Incoherent state & $\Delta_G[\mu] \in \mathcal{I}_G$ – Incoherent ensemble\\
 \hline
 Free operations & $\Lambda \in \mathcal{F}$ – Maximally Incoherent& $\mathcal{K} = \mathcal{K}_{P} \cdot \mathcal{K}_{\Phi} \in \mathcal{F}_G$ – Geometrically Incoherent\\
 \hline
\end{tabular}
}
\caption{Short summary that translates concepts from Quantum Mechanics (QM column) into the geometric approach (GQM column). Statistical states are density matrices in QM, while they are Geometric Quantum States in GQM, corresponding to ensembles of pure states. General transformations between states are described by CPTP maps in QM, while they are generic conditional probabilities–i.e. channels in information-theoretic language. At the measure-theoretic level, these are related to Radon-Nikodym derivatives. The channel $\Delta_G$ is the geometric equivalent to the standard dephasing channel $\Delta$ in that each channel erases the respective coherence from a given state. $\mathcal{I}$ and $\mathcal{F}$ represent incoherent states and operations at the density matrix level. Analogously, $\mathcal{I}_G$ and $\mathcal{F}_G$ identify incoherent geometric quantum states and the set of geometrically incoherent operations.}
\end{table*}
This motivates our central question: does $I(P;\Phi)$ satisfy the formal requirements of a coherence measure? In section \ref{sec:prob-phase-mutual-info-as-coherence}, we establish this by proving that $I(P;\Phi)$ satisfies six axioms, from resource theory \cite{Streltsov2017QuantumCoherenceResource}, that qualify it as a measure of coherence for ensembles of pure states.

\subsection{Experimental accessibility}

A natural practical question is what kind of access to the ensemble is required to extract $I(P;\Phi)$. Access to the density matrix $\rho$ alone is insufficient, as ensembles with arbitrarily different probability-phase correlations can produce the same density matrix, meaning standard state tomography is blind to ensemble-level coherence. Nevertheless, the density matrix does constrain $I(P;\Phi)$ via the coherence surplus bound (Theorem~\ref{eq:coherence-surplus}). If we instead consider samples of pure states drawn from the ensemble $\mu_{P,\Phi}$, the probability-phase mutual information can be estimated by discretizing the state space, building empirical histograms, and applying the scaling analysis of Eq. \eqref{eq:mutual_information_scaling} as detailed in section \ref{sec:examples}, without requiring analytical knowledge of the ensemble distribution. Whether obtaining such samples from the ensemble requires tomography is dependent on the physical setting. In the projected ensemble context (section \ref{sec:deep-thermalization}), each measurement on a subsystem $B$ directly yields a conditional pure state of $A$, from which ensemble samples can be obtained via conditional tomography of $A$.

Such ensemble sampling is no longer a theoretical abstraction: projected ensembles have been directly accessed in recent experiments on multiple platforms. Choi \emph{et al.}~\cite{choiPreparingRandomStates2023} demonstrated the emergence of random state ensembles from time-independent Hamiltonian dynamics using a Rydberg atom quantum simulator with up to 25 atoms. In their setup, the pure-state ensemble on subsystem $A$ is generated by performing projective measurements on the complement $B$, yielding conditional pure states whose distribution converges to the Haar-random ensemble. Importantly, their protocol requires no explicit time-resolved control, relying solely on the natural scrambling dynamics of the many-body system. More recently, Zhang \emph{et al.}~\cite{Zhang2025HolographicDeepThermalization} introduced holographic deep thermalization, a hardware-efficient scheme implemented on IBM Quantum devices that trades space for time by iterating scramble-measure-reset cycles with a constant number of ancilla qubits, achieving genuine random state generation for systems of up to 5 data qubits using only 8 qubits in total. These experiments demonstrate that ensemble-level observables---and hence quantities like $I(P;\Phi)$---are within reach of current quantum simulation platforms.

\section{Resource theory of coherence for ensembles of pure states}
\label{sec:prob-phase-mutual-info-as-coherence}

Resource theories~\cite{Chitambar2019QuantumResourceTheories} provide a rigorous framework for quantifying physical resources that cannot be freely created under a specified set of constraints. The resource theory of coherence, developed extensively in recent years~\cite{Baumgratz2014QuantifyingCoherence,Levi2014QuantitativeTheoryCoherent,Chitambar2016ComparisonIncoherentOperations,Chitambar2016CriticalExaminationIncoherent,Chitambar2017ErratumComparisonIncoherent,Winter2016OperationResourceTheory,Yadin2016QuantumProcessesCoherence}, identifies quantum coherence as a resource relative to a specific basis $\left\{ \ket{e_k}\right\}_{k=0}^{D-1}$. In the standard formulation, a density matrix $\rho$ is considered incoherent if it is diagonal in the reference basis $\rho_{kj}=\bra{e_k}\rho\ket{e_j}=\rho_{kk}\delta_{jk}$. The set of all incoherent density matrices is called $\mathcal{I}$. Incoherent operations are CPTP maps $\Lambda$ that cannot create coherence from incoherent states, thus preserving $\mathcal{I}$: $\mathcal{F}:= \left\{ \rho \in \mathcal{I} \Rightarrow \Lambda[\rho] \in \mathcal{I}\right\}$. Within this context, an important operation is \emph{dephasing}: $\Delta[\rho] = \sum_{k=0}^{D-1}\rho_{kk}\ket{e_k}\bra{e_k}$. This takes any density matrix $\rho$ and it extracts its incoherent part $\Delta[\rho]$. An important measure of coherence that emerges from this approach is the \emph{relative entropy of coherence}: $\mathcal{C}(\rho):=D^Q_{KL}(\rho \vert \vert \Delta[\rho])=S(\Delta[\rho]) - S(\rho)$, where $D_{KL}^Q(\rho || \sigma)$ is the quantum relative entropy between two density matrices and $S(\rho)$ is the von Neumann entropy.
\subsection{A geometric approach to quantum coherence in ensembles}
We leverage the geometric setting to systematically extend the resource theory of coherence to ensembles of pure states. This requires defining three new notions: \emph{geometric dephasing}, \emph{incoherent ensembles} and \emph{free operations}. 

In geometric quantum mechanics, the density matrix is replaced by the geometric quantum state---an ensemble of pure states described by the measure $\mu_{P,\Phi}$. Additionally, quantum operations are replaced by geometric channels $\mathcal{K}$: conditional probability measures $\mathcal{K}(Z|\Gamma)$. This is equivalent to the definition of channels in classical information theory. Specifically, a geometric channel $\mathcal{K}$ is a linear operation mapping a generic probability measure $\mu_n$ on $\mathbb{C}P^n$ into another probability measure $\sigma_m$ on $\mathbb{C}P^m$: $\mathcal{K}[\mu_n] = \sigma_m$. 

When written in terms of densities $d \mu_n = q_n(\Gamma) dV_{FS}^{\Gamma}$ and $d\sigma_m = \tilde{q}_m(Z)dV_{FS}^Z$ this takes the familiar form of a kernel, or conditional probability density:
\begin{align}
    \tilde{q}_m(Z) = \int_{\mathbb{C}P^{n}}\!\!\! dV_{FS}^\Gamma~~\mathcal{K}(Z \vert \Gamma) q_n(\Gamma) \nonumber
\end{align}
We now have everything we need to build the three core elements of our resource theory of coherence for ensembles. First, we define \emph{geometric dephasing} $\Delta_G$, as the operation which maps any geometric quantum state $\mu_{P,\Phi}$ into the product of its probability and phases marginals:
\begin{equation}
    \Delta_G[\mu_{P,\Phi}] = \mu_{P}~\cdot~\mu_{\Phi} \nonumber
\end{equation}
This translation shifts the focus from coherence as superposition within individual quantum states to coherence as statistical correlation within quantum ensembles.

Paralleling the standard resource theory of quantum coherence, $\mu_{P,\Phi}$ is considered incoherent when it has a form compatible with the output of the dephasing channel $\Delta_G$. That is, when probability and phase variables are statistically independent $\mu_{P,\Phi} = \mu_P \cdot \mu_\Phi$. This condition ensures that knowledge of probabilities provides no information about the phases, and vice versa. The set of \emph{incoherent ensembles} is thus:
\begin{align}
\mathcal{I}_G = \{\mu_P \cdot \mu_\Phi \}~, \label{eq:free_states}
\end{align}
where $\mu_P$ is a measure on $\Sigma_{D-1}$ and $\mu_\Phi$ is a measure on $\mathbb{T}_{D-1}$.

Free operations in this resource theory are those that preserve the factorized structure of incoherent states. Specifically, a channel $\mathcal{K}$ acting on geometric quantum states is free if it can be decomposed as $\mathcal{K} = \mathcal{K}_P \cdot \mathcal{K}_\Phi$, where $\mathcal{K}_P$ acts only on the probability variables and $\mathcal{K}_\Phi$ acts only on the phase variables. The set of \emph{free operations} is therefore:
\begin{align}
\mathcal{F}_G = \{\mathcal{K}_P \cdot \mathcal{K}_\Phi\} \nonumber
\end{align}
The choice of factorized channels is physically motivated: it is the structure required to ensure that all free states (product measures) remain product measures. To see why, consider the simplest operation that does not factorize: $\mathcal{K} = \sum_i q_i\mathcal{K}_P^{i}\cdot\mathcal{K}_\Phi^i$. Acting on a product measure $\mu = \mu_P\cdot\mu_\Phi$ produces a state $\tilde{\mu} = \sum_i q_i \mu_P^i\cdot\mu_\Phi^i$ which, in general, will not be incoherent $\tilde{\mu} \notin \mathcal{I}_G$. This is a reflection of the fact that we can create a correlated distribution of two random variables by mixing different product distributions: $p_{corr}(x,y)=\alpha p_1(x)p_1(y)+(1-\alpha)p_2(x)p_2(y)$. Factorized channels are therefore the physically natural class of free operations within this framework.


With this setup, we now apply the framework established by Streltsov, Adesso, and Plenio~\cite{Streltsov2017QuantumCoherenceResource}. For $I(P;\Phi)$ to qualify as a coherence measure, it must satisfy the following six properties. We now explicitly state the properties, and comment on their physical relevance in our resource theory.

\textbf{(C1) Non-negativity:} $I(P;\Phi) \geq 0$ with equality if and only if $\mu_{P,\Phi} \in \mathcal{I}_G$. This follows directly from the non-negativity of mutual information, see Eq.(\ref{eq:mutual_information_relative_entropy}), and the fact that $D_{KL}(\mu||\nu) = 0$ if and only if $\mu = \nu$.

\textbf{(C2) Monotonicity:} $I(P;\Phi)$ does not increase under free operations. For any $\mathcal{K} \in \mathcal{F}_G$, we have $I(\mathcal{K}[\mu_{P,\Phi}]) \leq I(\mu_{P,\Phi})$. This guarantees that when we act with free operations resources can only be depleted, not created.

\textbf{(C3) Strong monotonicity:} $I(P;\Phi)$ does not increase on average under selective free operations. This property guarantees that measurements do not create ensemble coherence.

\textbf{(C4) Convexity:} $I(P;\Phi)$ is a convex function of the geometric quantum state, guaranteeing that classically mixing two ensembles can only deplete coherence.

\textbf{(C5) Uniqueness for pure states:} For pure states, described by a Dirac measure with support on a single point $\mu^{\mathrm{pure}}_{P,\Phi} = \delta_{p_0,\phi_0}$, the probability-phase mutual information vanishes:
\begin{equation}
    I(P;\Phi) = H(\Delta_G[\delta_{p_0,\phi_0}]) = 0 \nonumber
\end{equation}
This property distinguishes $I(P;\Phi)$ fundamentally from density-matrix coherence measures. A pure state $\ket{\psi}$ in superposition can have $\mathcal{C}(\rho) > 0$, yet when viewed as an ensemble (a single point in state space), it necessarily has $I(P;\Phi) = 0$. The probability-phase mutual information therefore captures coherence that emerges only at the ensemble level—it quantifies correlations across a statistical distribution of pure states, not superposition within individual states. This is the additional resource that is invisible to density matrices and that becomes relevant for when density matrices are mixed rather than pure.

\textbf{(C6) Additivity:} For non-interacting systems described by $\mu_{P,\Phi} \cdot \nu_{P',\Phi'}$, we have $I(P,P';\Phi,\Phi') = I(P;\Phi) + I(P';\Phi')$. This guarantees that we do not create our resource, mutual information, by simply putting together non-interacting systems.

Detailed proofs that $I(P,\Phi)$ satisfies all these properties are given in Appendix \ref{appendix}. The net result establishes $I(P;\Phi)$ as an operationally meaningful measure of coherence for ensembles of pure states. A discussion to compare it with density-matrix-based measures of coherence, like the relative entropy of coherence \cite{Streltsov2017QuantumCoherenceResource,herbut_quantum_2005}, is given in Section \ref{sec:discussion}. 

\section{Examples}
\label{sec:examples}

We have concluded the technical development of the probability-phase mutual information, and established its validity as a coherence measure for ensembles. The next subsections show how to compute $I(P,\Phi)$, both analytically and numerically, in a variety of cases: Product measures, Canonical ensemble, Gaussian measure, Spiral ensemble, and Deep thermalization. 
These illustrate a wide variety of probability-phase correlations that can arise in geometric quantum states, from the complete absence of correlations to maximal correlations in coherent ensembles. 

For geometric quantum states where analytical calculations are intractable, we employ a direct numerical approach that leverages the discrete partitioning of complex projective spaces detailed in Ref.\cite{Anza2022QuantumInformationDimension}:

\begin{enumerate}
    \item Discretization: Partition the state space into cells of linear size $\epsilon$ 
    \item Histogram estimation: Coarse grain $\mu_{P,\Phi}$ to compute empirical probabilities of the joint and marginals
    \item Entropy calculation: Apply the scaling analysis to extract $\mathfrak{D}_{P\Phi},\mathfrak{D}_P,\mathfrak{D}_\Phi,H_{P\Phi},H_P,H_\Phi$
    \item Mutual information: Compute via Eq.  \eqref{eq:mutual_information_scaling}: $I(P;\Phi) = H_P + H_\Phi - H_{P\Phi}$
\end{enumerate}

This procedure has been validated against the analytical examples presented, showing convergence as $\epsilon \to 0$ with the expected scaling behavior.

\subsection{Example 1: Product Measures}\label{subsec:product-measures}
We first consider a class of ensembles which share the common property that they can be written as product measures.

\paragraph{Dirac measure.} Consider an ensemble composed of a single pure state, corresponding to a concentration of measure on one point in the quantum state space. This is represented by the Dirac measure: $\delta^{\Psi}_{\psi_0} = \delta^{P}_{p_0} \delta^{\Phi}_{\phi_0}$ where $\delta_{p_0}^P$ is the Dirac measure on $\Sigma_{D-1}$ with support on $p_0$ and  $\delta_{\phi_0}^{\Phi}$ is the Dirac measure on $\mathbb{T}_{D-1}$ with support on $\phi_0$. The joint measure is clearly a product of its marginals, so the mutual information vanishes.

\paragraph{Uniform measure.} The uniform (Haar) ensemble is also a product measure. For this ensemble, every point in $\mathcal{P}(\mathcal{H}) \sim \mathbb{C}P^{D-1}$ is equally probable. For a $D$-dimensional Hilbert space the uniform distribution is written in terms of the Fubini-Study volume $V_{D-1}=\frac{\pi^{D-1}}{(D-1)!}$ \cite{Anza2022QuantumInformationDimension}. Using Eq. \eqref{eq:fubini-study-element} and the Fubini-Study volume, the uniform density in probability-phase coordinates is:
\begin{equation}
    q_{P,\Phi}(p,\phi) = \frac{(D-1)!}{(2\pi)^{D-1}} \nonumber
\end{equation}

Integration over the probability simplex $\Sigma_{D-1}$ and hyper-torus $\mathbb{T}_{D-1}$ yields the marginals $q_P(p) = (D-1)!$ and $q_\Phi(\phi) = \frac{1}{(2\pi)^{D-1}}$, confirming the factorization and vanishing of mutual information.

\paragraph{Diagonal ensemble.} Given some isolated system initialized in $\ket{\psi_0}$ and evolving with hamiltonian $H = \sum_n E_n \ket{E_n} \bra{E_n}$ with non-degenerate energy gaps, the distribution of coefficients $c_n(t)=\braket{E_n \vert \psi_t}=|c_n(0)|e^{-\frac{i}{\hbar}E_nt}$ resulting from time-evolution is, as obtained by von Neumann \cite{goldstein_long-time_2010}, the product of a Dirac measure in $\Sigma_{D-1}$ and a uniform measure in phases:
\begin{equation}
    \mu_{DE} = \delta^P_{p_0} \cdot \frac{1}{(2\pi)^{D-1}} \nonumber
\end{equation}
where $p_0$ fixes the probability coordinates in the energy basis to be $p_0 = (|c_1^2|,\ldots,|c_{D-1}|^2)$. This has vanishing probability-phase mutual information in the energy basis. We note that $\mu_{DE}$ is the ensemble whose average provides the stationary state commonly called \emph{diagonal ensemble}: $\mathbb{E}_{\mu_{DE}}[\psi] = \sum_{n}|c_n|^2 \ket{E_n}\bra{E_n}$.

\paragraph{Naive gaussian.} A final example of a product measure is the naive Gaussian measure:
\begin{equation}
    q(p,\phi) = \frac{1}{\mathcal{N}}\,\exp{\left[ -\frac{(p-p_0)^2}{2\sigma_p^2} - \frac{(\phi - \phi_0)^2}{2\sigma_\phi^2} \right]} \nonumber
\end{equation}
The marginal distributions are then gaussian in each coordinate with normalization constants satisfying $\mathcal{N} = \mathcal{N}_p\mathcal{N}_\phi$ which requires the joint measure be a product of marginals.

In each of these cases, the factorization of the joint distribution means that the joint  distribution is a product measure $\mu_{P,\Phi} = \mu_p\cdot\mu_\phi$. Therefore, from Eq. \eqref{eq:free_states}, $\mu_{P,\Phi} \in \mathcal{I}_G$ for each of these ensembles and $I(P,\Phi)=0$.

\subsection{Example 2: Canonical Ensemble}

\begin{figure}[t]
    \centering
    \includegraphics[scale=0.3]{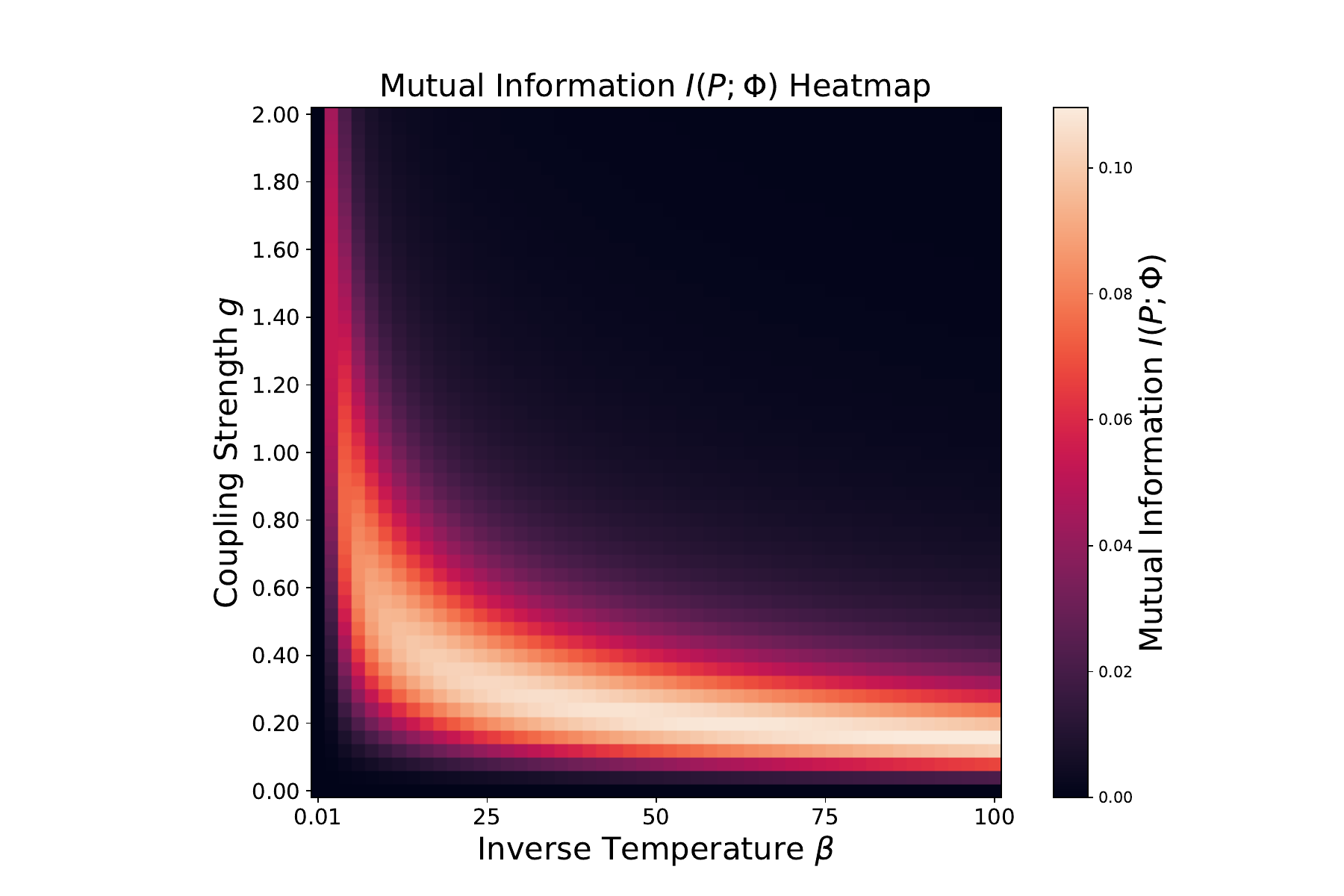}
    \caption{Probability–phase mutual information $I(P;\Phi)$ for the canonical ensemble of a qubit with Hamiltonian $H=\sigma_z+g\sigma_x$, shown as a function of inverse temperature $\beta$ and coupling strength $g$. As expected, $I(P;\Phi)$ vanishes when $g=0$. At non-vanishing coupling, correlations emerge: the mutual information grows in the intermediate temperature regime and peaks at small $g$, reflecting stronger probability–phase dependence in ordered regimes with nonzero temperature. Note that $I(P;\Phi)$ will approach zero as $\beta\to\infty$ since the canonical ensemble becomes pure in this limit. This implies that $I(P;\Phi)$ reaches a peak value at finite temperature.
}
    \label{fig:canonical_ensemble_numerics}
\end{figure}

Consider a canonical ensemble $q_\beta(p,\phi)$ at inverse temperature $\beta$ and with Hamiltonian $H(g)$ where $g$ is a coupling parameter:
\begin{subequations}
\begin{align}
    q_{\beta,g}(p,\phi) &= \frac{1}{Z_{\beta,g}}\exp{(-\beta h_g(p,\phi))} \nonumber \\
    Z_{\beta,g}&=\int_{\Sigma_{D-1}}\!\!\! dp \int_{\mathbb{T}_{D-1}} \!\!\!d \phi~e^{- \beta h_g(p,\phi)}\nonumber
    \label{eq:canonical_ensemble}
\end{align}
\end{subequations}
where $h_g(p,\phi) = \bra{p,\phi}H(g)\ket{p, \phi}$ is the expectation value of $H(g)$ on a generic pure state $\ket{p,\phi} = \sum_{j=1}^{D} \sqrt{p_j}e^{i\phi_j}\ket{e_j}$ parametrized with probability-phase coordinate in the basis $\left\{\ket{e_j}\right\}_j$ that diagonalizes $H(g=0)$: $H(g=0)\ket{e_j} = e_j \ket{e_j}$.

For a qubit with Hamiltonian $H = \sigma_z + g\sigma_x$, using the Pauli-z basis we have
\begin{equation}
    h_g(p,\phi) = 1 - 2p + 2g\sqrt{p(1-p)}\,\cos\phi \nonumber
\end{equation}
Notice that correlations between the probabilities and phases are introduced through the final term involving $\cos\phi$. This term results from the choice of the Pauli-z basis to represent $h_g(p,\phi)$, explicitly showing that the probability-phase correlations, and therefore $I(P;\Phi)$, are basis dependent. Choosing a different basis—for instance the eigenbasis of $H(g)$—would yield a different decomposition of $h_g(p,\phi)$ and, in general, a different value of $I(P;\Phi)$. This basis dependence is a standard feature shared by all coherence measures. To evaluate $I(P;\Phi)$, the marginal distributions are obtained by partial integration over the probability and phase degrees of freedom.

In the case of no coupling ($g=0$) these can be evaluated analytically: the joint distribution factorizes and the mutual information vanishes. In the general case $g > 0$, the joint distribution does not factorize. We evaluate $I(P;\Phi)$ numerically within the $\beta$-$g$ parameter space. The results are shown in Figure \ref{fig:canonical_ensemble_numerics}. For finite coupling $g > 0$, the behavior of $I(P;\Phi)$ as a function of $\beta$ is non-monotonic: it initially grows as the temperature decreases, reflecting increasing statistical dependence between probability and phase coordinates, but must eventually return to zero as $\beta \to \infty$ ($T\to 0$) since the canonical ensemble becomes a pure state in this limit. The probability-phase mutual information therefore peaks at a finite temperature that depends on $g$. At fixed $\beta$, $I(P;\Phi)$ also varies with the coupling strength $g$: it vanishes at $g=0$ (where the Hamiltonian is diagonal in the chosen basis) and grows as $g$ introduces off-diagonal structure that couples probabilities to phases. The dependence on $g$ is also non-monotonic: since the probability-phase coupling enters the canonical weight through the effective parameter $\beta g$, large $g$ at fixed $\beta$ concentrates the ensemble around the minimum of $h_g(p,\phi)$, ultimately approaching a Dirac measure with vanishing $I(P;\Phi)$.
We highlight that the rationale for the choice of the canonical ensemble goes beyond the need to illustrate the behavior of the probability-phase mutual information. Its relevance stems from the concept of conditional thermalization \cite{Sone2021JarzynskiEqualityConditional,Sone2024ConditionalQuantumThermometry}, geometric quantum thermodynamics \cite{Anza2022GeometricQuantumThermodynamics} and deep thermalization \cite{Ippoliti2022SolvableModelDeep, Mark2024MaximumEntropyDeep, Rigol2010QuantumChaosThermalization, Lucas2023GeneralizedDeepThermalization, Ippoliti2023DynamicalPurificationEmergence, Bhore2023DeepThermalizationConstrained, Ho2022ExactEmergentQuantum}. The discussion on this point is postponed to Section \ref{sec:deep-thermalization}.

\subsection{Example 3: Gaussian Measure}
\begin{figure}[ht]
    \centering
    \includegraphics[scale=0.5]{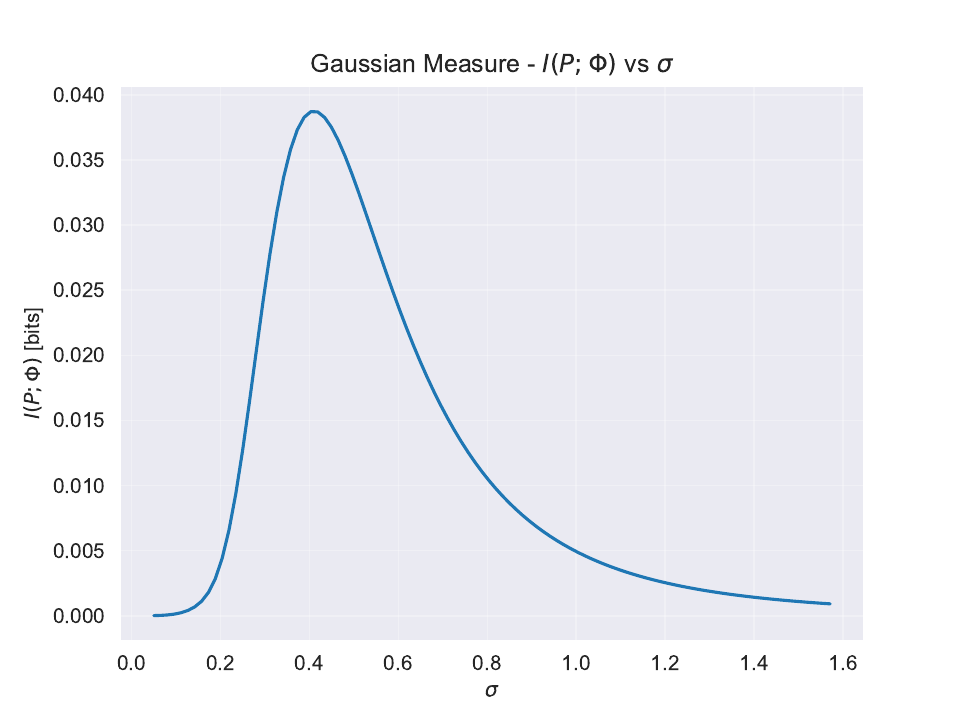}
    \caption{Probability–phase mutual information $I(P;\Phi)$ for a Gaussian ensemble defined by the Fubini–Study distance $D_{FS}(p,\phi; p_0, \phi_0)$ from a reference state $p_0=1/2$, $\phi_0 = \pi$ as a function of the width $\sigma$. For small $\sigma$, the ensemble is sharply localized (approaching a Dirac measure) and $I(P;\Phi)$ is near zero. Increasing $\sigma$ introduces nontrivial correlations between probabilities and phases, leading to a peak in mutual information before the ensemble approaches the uniform distribution at large $\sigma$, where $I(P;\Phi)\to 0$.}
    \label{fig:gaussian_mi_v_sigma}
\end{figure}
Consider a Gaussian density defined using the Fubini-Study distance $D_{FS}=D_{FS}(p,\phi;p_0,\phi_0)$ expressed in probability-phase coordinates. This density results from the extension of the standard Gaussian density on metric manifolds, intended as exponentials of squared distances. The distribution can be written as:
\begin{equation}
    \mathcal{G}_2(p,\phi) = \frac{1}{\mathcal{N}}\exp{\left[ -\frac{D_{FS}(p,\phi;p_0,\phi_0)^2}{2\sigma^2} \right]} \nonumber
\end{equation}
where the Fubini-Study distance for a single qubit is: $D_{FS}(p,\phi;p_0,\phi_0) = \arccos{|\braket{\psi(p,\phi)|\psi(p_0,\phi_0)}|}$, with $\bra{\psi(p,\phi)}\psi(p_0,\phi_0)\rangle = \sqrt{(1-p_0)(1-p)} + \sqrt{p_0p }~e^{i(\phi_0-\phi)}$

Since $D_{FS}$ does not factorize as a sum over $p$ and $\phi$, it introduces non-trivial probability-phase correlations. As shown in Figure \ref{fig:gaussian_mi_v_sigma}, $I(P;\Phi)$ is non-monotonic in $\sigma$: at $\sigma = 0$ the ensemble is a Dirac measure with vanishing mutual information; as $\sigma$ grows, correlations build up and $I(P;\Phi)$ reaches a maximum; for $\sigma \gg \max_{p,\phi}D_{FS} = \pi/2$ the ensemble approaches the uniform (Haar) measure and $I(P;\Phi)$ returns to zero.

\subsection{Example 4: Ensemble Thermalization}
\label{sec:deep-thermalization}

In quantum thermalization the density matrix of an open quantum system relaxes to a steady state described by a thermal state, e.g, Gibbs' canonical density matrix. Instead of framing the question at the density matrix level one can talk about \emph{ensemble thermalization}, referring to the idea that the ensemble behind a density matrix acquires a thermal equilibrium statistics. This phenomenon has been studied under three different names. It is known with ``Deep Thermalization'' \cite{Ippoliti2022SolvableModelDeep, Mark2024MaximumEntropyDeep, Rigol2010QuantumChaosThermalization, Lucas2023GeneralizedDeepThermalization, Ippoliti2023DynamicalPurificationEmergence, Bhore2023DeepThermalizationConstrained, Ho2022ExactEmergentQuantum} within the condensed matter and quantum information community. And, it has been investigated under the name of ``Universal Probability Distribution'' \cite{Goldstein2016UniversalProbabilityDistribution}, ``Geometric thermalization'' \cite{Anza2022GeometricQuantumThermodynamics} or ``Conditional Thermalization'' \cite{Sone2021JarzynskiEqualityConditional, Sone2024ConditionalQuantumThermometry} in Quantum Thermodynamics. While framed under slightly different lights, the core idea behind all these advanced studies of quantum thermalization and the resulting thermodynamic behavior is the same.

We have a unitarily evolving bipartite quantum system in a pure state: $\ket{\Psi_t} \in \mathcal{H}_A \otimes \mathcal{H}_B$. Measuring the complement $B$ in an orthonormal basis $\{\ket{z_B}\}_{z_B}$ and collecting the conditional pure states of $A$,
\begin{subequations}
\begin{align}
  p_t(z_B) &= \braket{\Psi_t | (\mathbb{I}_A\cdot \ket{z_B}\!\bra{z_B}) |\Psi_t}, \nonumber \\
  \ket{\Psi_t(z_B)} & = \frac{(\mathbb{I}_A\cdot \bra{z_B})\ket{\Psi_t}}{\sqrt{p_t(z_B)}}, \nonumber
\end{align}
\end{subequations}
leads to the \emph{projected ensemble} $\mathcal{E}_{\mathrm{proj}}(t)=\{p_t(z_B),\,\ket{\Psi_t(z_B)}\}_{z_B}$ on $A$. This is a conditional ensemble and also the geometric quantum state on $\mathcal{P}(\mathcal{H}_A)$. Deep thermalization posits that \cite{Ho2022ExactEmergentQuantum}, in chaotic dynamics and at very high energy density (no conservation constraints), the geometric quantum state approaches the unitarily invariant (Haar) distribution on $A$—i.e., an \emph{infinite-temperature uniform ensemble}. 

In our geometric formalism, this simply means constant density with respect to the Fubini–Study (FS) volume element. As shown in Sec. \ref{subsec:product-measures}, this leads to a \emph{vanishing  mutual information}. This gives a sharp, basis-resolved statement: deep thermalization $\Rightarrow$ $I(P,\Phi)=0$, no statistical dependence between probabilities and phases at the ensemble level. Viceversa, a non-vanishing correlation between probabilities and phases $I(P,\Phi) \neq 0$ is sufficient to rule out deep thermalization. 

To test this prediction numerically, we consider a linear chain of $N$ spin-1/2 systems interacting via the quantum Ising model with mixed fields (QIMF):
\begin{align}
    H_{\mathrm{QIMF}} = J\sum_{i=1}^{N-1}\sigma^x_i\sigma^x_{i+1} + h\sum_{i=1}^{N}\sigma^y_i + g\sum_{i=1}^{N}\sigma^x_i \nonumber
\end{align}
with parameters $J=1$, $h=0.9045$, $g=0.8090$. This model is known to be chaotic and its projected ensembles have been shown to form approximate quantum state $k$-designs~\cite{cotlerEmergentQuantumState2023}, making it an ideal testbed. Our subsystem of interest $A$ is a single qubit located at the center of the spin chain, while the environment $B$ consists of the remaining $N-1$ spins measured in the computational basis $\ket{z_B} = \ket{s_1,\ldots,s_{N-1}}$, with $s_i \in \left\{ 0,1\right\}$.

Starting from an initial product state $\ket{\Psi_0} = \ket{0}^{\otimes N}$, we evolve the system under $H_{\mathrm{QIMF}}$ and construct the projected ensemble at each time $t$. To mitigate finite-size effects and transient short-time dynamics, we consider the \emph{time-averaged} projected ensemble, obtained by averaging the empirical distribution of conditional pure states over a sufficiently long time window. This increases the effective statistics of the ensemble and suppresses fluctuations that, at finite system size, can spuriously inflate $I(P;\Phi)$.

\begin{figure}[h!]
    \centering
    \includegraphics[scale=0.5]{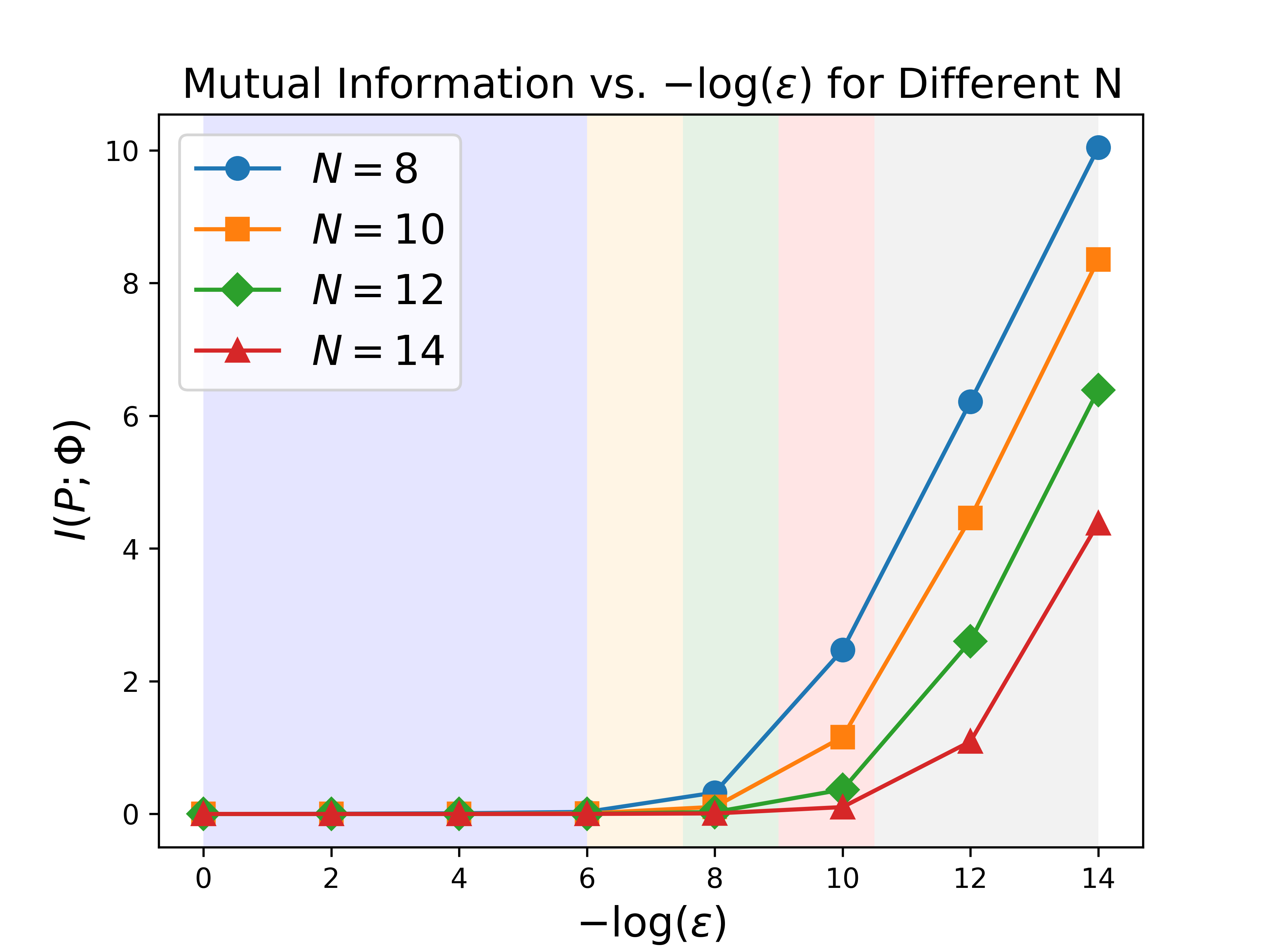}
    \caption{Scaling analysis of the discretized mutual information $I_\epsilon(P;\Phi)$ as a function of $-\log\epsilon$ for the time-averaged projected ensemble of the QIMF model, for system sizes $N=8,10, 12,14$. The blue, orange, green and red shaded regions identify the plateau where $\mathfrak{D}_I = 0$ and $I_\epsilon(P;\Phi) = 0$ for different system size. Larger system sizes exhibit a plateau that extends at finer scales (smaller $\epsilon$), thus confirming the results in Ref.\cite{cotlerEmergentQuantumState2023} arguing in favor of the emergence of deep thermalization in the thermodynamic limit.}
    \label{fig:deep-thermalization}
\end{figure}

We then apply the scaling analysis of Eq.~\eqref{eq:mutual_information_scaling} to the time-averaged ensemble. The results are shown in Figure~\ref{fig:deep-thermalization} for system sizes $N=8,10,12,14$. At coarse resolutions (small $-\log\epsilon$), the discretized mutual information $I_\epsilon(P;\Phi)$ remains near zero, exhibiting a plateau consistent with a vanishing mutual information dimension $\mathfrak{D}_I=0$ and, correspondingly, a vanishing probability-phase mutual information $I(P;\Phi)=0$. This is the scaling signature expected when the joint and marginal supports coincide, as is the case for the Haar distribution. At finer resolutions (large $-\log\epsilon$), $I_\epsilon(P;\Phi)$ grows due to finite-size effects: the limited number of conditional pure states in the projected ensemble ($2^{N-1}$ outcomes) cannot adequately resolve the fine-grained structure of the distribution, producing spurious correlations. Crucially, the extent of the $\mathfrak{D}_I=0$ plateau grows systematically with system size---larger $N$ pushes the onset of finite-size artifacts to finer scales. This behavior strongly suggests that, in the thermodynamic limit $N\to\infty$, the plateau extends over the entire range of resolutions, confirming that the probability-phase mutual information vanishes: $I(P;\Phi)=0$. Since a vanishing $I(P;\Phi)$ in the Pauli-$z$ basis is the signature of deep thermalization in our framework, these results provide direct numerical evidence that supports the conclusions in Ref. \cite{cotlerEmergentQuantumState2023} that the projected ensemble of the QIMF model deeply thermalizes.

\section{Discussion}
\label{sec:discussion}

The probability–phase mutual information $I(P;\Phi)$ and the relative entropy of coherence $\mathcal{C}(\rho)$ quantify fundamentally different types of coherence that emerge at ensemble and density matrix levels. While $\mathcal{C}(\rho)$ quantifies coherence by evaluating the relevance of the off-diagonal matrix elements in a given basis, $I(P;\Phi)$ does it by addressing the question of how much information about phases is accessible from the probabilities. 

Crucially, this distinction is sharp for pure states: a $\ket{\psi}$ in superposition has $\mathcal{C}(\ket{\psi}\bra{\psi}) > 0$, yet when viewed as an ensemble—a Dirac measure concentrated at a single point—necessarily has $I(P;\Phi) = 0$. No spread means no correlations. Therefore, $I(P,\Phi)$ \emph{captures a form of coherence that is available only to mixed states}. 

This raises a natural question: when a mixed state exhibits both types of coherence, how do they relate? The same density matrix $\rho$ can arise from ensembles with different correlation structures, so $I(P;\Phi)$ and $\mathcal{C}(\rho)$ can vary independently. Yet, they are related. The \emph{coherence surplus}, which we define in the next subsection, establishes a quantitative connection between them, revealing how ensemble correlations contribute to density-matrix coherence. 


\subsection{Relating ensemble and density-matrix coherence via the coherence surplus}

\begin{figure*}[ht]
\centering
\includegraphics[width=0.8\textwidth]{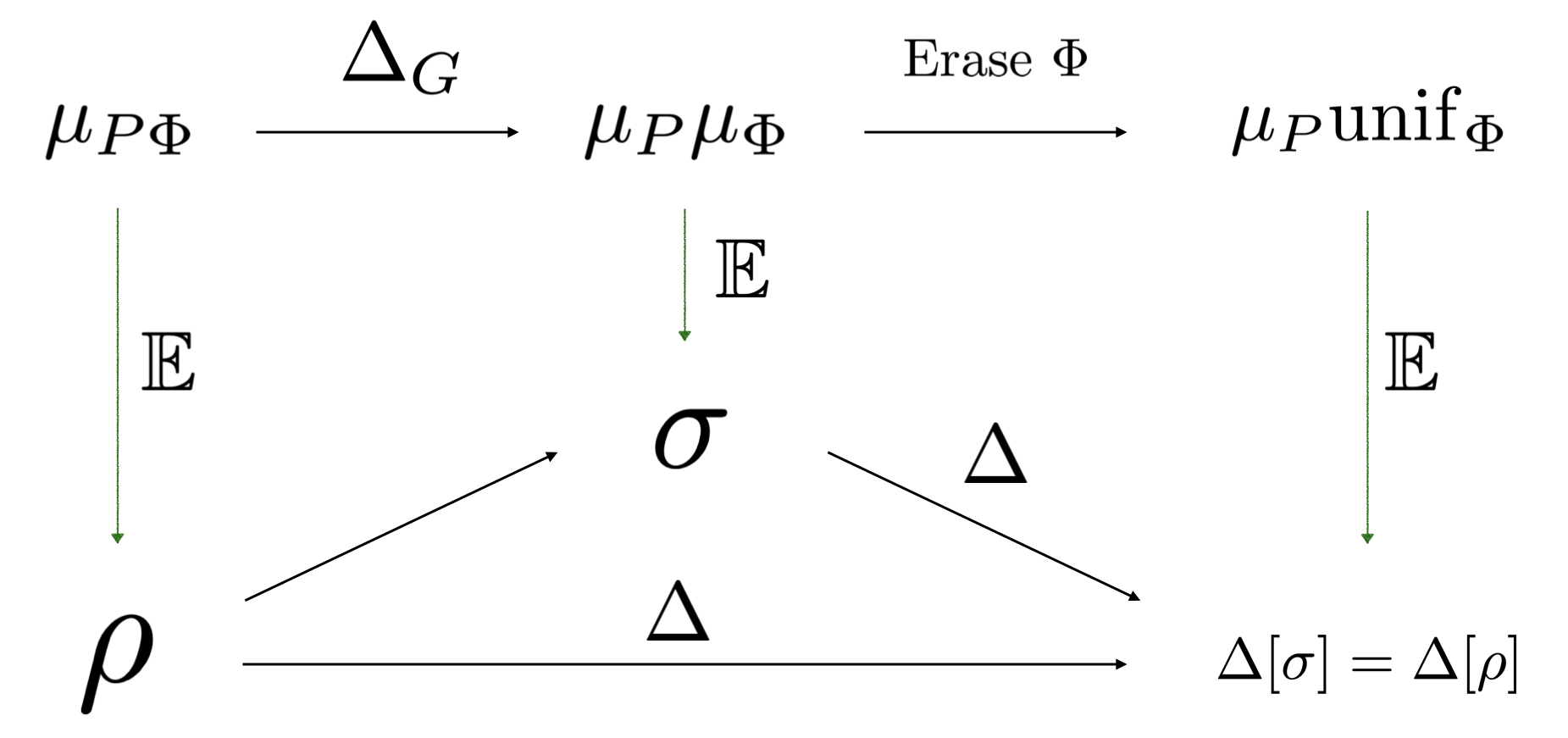}
\caption{Commutative diagram relating ensemble and density-matrix coherence. The top row shows transformations at the ensemble level: geometric dephasing $\Delta_G$ factorizes the joint probability-phase measure, followed by phase uniformization. The bottom row shows the corresponding density matrix transformations, where $\sigma$ is the intermediate state after geometric dephasing. The vertical arrows represent the expectation operator $\mathbb{E}$ mapping ensembles to their density matrix. The diagram commutes at both ends, establishing the connection between probability-phase mutual information $I(P;\Phi)$ and the relative entropy of coherence $\mathcal{C}(\rho)$.}
\label{fig:coherence_diagram}
\end{figure*}

The relation between $I(P,\Phi)$ and $\mathcal{C}(\rho)$ becomes clear through the diagram in Figure~\ref{fig:coherence_diagram}, which illustrates a two-step transformation process. The top row depicts the ensemble-level transformations. Starting from a joint probability-phase measure $\mu_{P,\Phi}$, geometric dephasing $\Delta_G$ factorizes it into the product of marginals $\mu_P \cdot \mu_\Phi$. A subsequent phase uniformization replaces the phase marginal with the uniform distribution, yielding $\mu_P \cdot \mathrm{unif}_\Phi$. The bottom row shows the corresponding density-matrix transformations, connected by the expectation operator $\mathbb{E}$ that maps ensembles to their density matrices:
\begin{align*}
\rho & = \mathbb{E}_{\mu_{P\Phi}}[\psi] \\
\sigma &= \mathbb{E}_{\mu_P \cdot \mu_\Phi}[\psi]\\
\Delta[\rho] & = \mathbb{E}_{\mu_P \cdot \mathrm{unif}_\Phi}[\psi] =  \Delta[\sigma] 
\end{align*}

At the ensemble level, the relative entropy $D_{\mathrm{KL}}(\mu_{P\Phi} \| \mu_P \cdot \mathrm{unif}_\Phi)$ quantifies how far the geometric quantum state is from being both incoherent and with uniform distribution of phases. At the density-matrix level, the Kullback-Leibler divergence $\mathcal{C}(\rho) = D^Q_{\mathrm{KL}}(\rho \| \Delta[\rho])$ measures the relative entropy between $\rho$ and the dephased state $\Delta[\rho]$. The difference between these two levels—the coherence surplus—captures how much coherence structure the ensemble contains beyond what is visible in the density matrix:

\begin{definition}[Coherence Surplus]\label{eq:coherence-surplus-definition} Let $\mu_{P\Phi}$ be a geometric quantum state on $\mathbb{C}P^{D-1}$ with probability and phase marginals $\mu_P$ and $\mu_\Phi$, respectively. Let $\rho = \mathbb{E}_{\mu_{P,\Phi}}[\psi]$ be its associated density matrix, and $\mathrm{unif}_\Phi$ denote the uniform phase distribution on $\mathbb{T}_{D-1}$. We define the coherence surplus as
\begin{align}
    \delta_{\mathcal{C}} := D_{\mathrm{KL}}(\mu_{P\Phi} \| \mu_P \cdot \mathrm{unif}_\Phi) - D^Q_{\mathrm{KL}}(\rho \| \Delta[\rho])
\end{align}
\end{definition}
Since the ensemble-level relative entropy decomposes as $D_{\mathrm{KL}}(\mu_{P\Phi} \| \mu_P \cdot \mathrm{unif}_\Phi) = I(P;\Phi) + D_{\mathrm{KL}}(\mu_\Phi \| \mathrm{unif}_\Phi)$ via chain rule, we obtain the following theorem:
\begin{theorem}\label{eq:coherence-surplus}
The coherence surplus, as per Definition \ref{eq:coherence-surplus-definition}, is always non-negative 
\begin{align}
\delta_{\mathcal{C}} =  I(P;\Phi) + D_{\mathrm{KL}}(\mu_\Phi \| \mathrm{unif}_\Phi) - \mathcal{C}(\rho) \geq 0 \nonumber
\end{align}
\end{theorem}

The proof, provided in Appendix~\ref{apx:coherence-theorem}, relies on the commutativity of the diagram in Figure~\ref{fig:coherence_diagram} and the quantum data-processing inequality.

The positivity of $\delta_{\mathcal{C}}$ quantifies the coherence discrepancy between the ensemble and density-matrix level of description. We highlight three physical consequences that descend from this, which are of immediate interest.

\textit{Coherence bound.} The positivity of $\delta_{\mathcal{C}}$ immediately yields a novel upper bound to the relative entropy of coherence:
\begin{equation}
\mathcal{C}(\rho) \leq I(P;\Phi) + D_{\mathrm{KL}}(\mu_\Phi \| \mathrm{unif}_\Phi)
\label{eq:global-bound}
\end{equation}
This reveals the fundamental asymmetry between ensembles and density matrices: the ensemble always contains at least as much coherence structure as the density matrix. This is physically intuitive—the geometric quantum state captures the full distribution of pure states, while the density matrix retains only their average. Correlations visible at the ensemble level may be washed out by averaging, but cannot be created by it. The coherence surplus quantifies this information loss.

\textit{Classical-quantum decomposition.} The coherence surplus naturally identifies to independent contributions to coherence. The mutual information $I(P;\Phi)$ characterizes how much phase information is accessible from observing the probabilities, classically accessible: even though phases themselves cannot be directly measured in the reference basis, their statistical relationship to probabilities creates an indirect information channel. In contrast, the ``phase non-uniformity'' $D_{\mathrm{KL}}(\mu_\Phi||\mathrm{unif}_\Phi)$ characterizes phase structure that remains completely inaccessible to any observer restricted to the reference basis, reflecting the non-commutative nature of quantum measurements. An ensemble that lacks both must necessarily produce a diagonal density matrix.

\textit{Entropy Gap Bound.} Recalling Definition \ref{eq:coherence-surplus-definition} and expanding the relative entropies as entropy differences, the coherence surplus leads to a counterintuitive bound between the two entropy gaps:
\begin{align}
    H_G(\mu_P\mathrm{unif}_\Phi) - S(\Delta[\rho]) \ge H_G(\mu_{P\Phi}) - S(\rho)~. \nonumber
\end{align}
Since the von Neumann entropy is the entropy of the eigenensemble of $\rho$, the quantities on both sides of these inequality are positive, as proven in \cite{Anza2022QuantumInformationDimension,Anza2024MaximumGeometricEntropy}. However, the existence of a specific order between them is non-trivial and it is due to the non-negativity of $\delta_{\mathcal{C}}$. The bound indicates that geometric dephasing $\Delta_G$––operating at the ensemble level––adds more classical noise to $\mu_{P\Phi}$ than the standard dephasing $\Delta$ does to the density matrix $\rho$, thus increasing the gap between geometric and von Neumann entropy.

\subsection{Operational Consequence of $I(P;\Phi)$}

In previous sections, we establish $I(P;\Phi)$ as a formally valid coherence measure for ensembles and relate it to the relative entropy of coherence $\mathcal{C}(\rho)$ through the coherence surplus. A natural question remains: beyond its formal properties, what physical content or operational consequences does $I(P;\Phi)$ possess? We address this question by first discussing an operational consequence of the relative entropy of coherence $\mathcal{C}(\rho)$. In the standard resource theory of coherence, it has been shown that the probability $P(\rho\to\sigma)$ of being able to convert a state described by the density matrix $\rho$ to a state with density matrix $\sigma$ via stochastic incoherent operations (SIO) is bounded by \cite{Wu2020QuantumCoherenceState, Du2015CoherenceMeasuresOptimal}:
\begin{equation}
    P(\rho\to\sigma) \le \frac{\mathcal{C}(\rho)}{\mathcal{C}(\sigma)} \notag
\end{equation}
where $\mathcal{C}$ is any measure of coherence satisfying strong monotonicity (C3) and non-negativity (C1). Since $I(P;\Phi)$ satisfies both of these axioms, this argument extends identically to the ensemble level. 

Let $\mu$ and $\nu$ be geometric quantum states and let $I_\xi(P;\Phi)$ be the probability-phase mutual information associated with the geometric quantum state $\xi$. Since $I(P;\Phi)$ satisfies axioms (C1) and (C3), the probability of conversion from $\mu$ to $\nu$ via stochastic free operations is bounded by:
\begin{equation}
    P(\mu\to\nu) \le \frac{I_\mu(P;\Phi)}{I_\nu(P;\Phi)} \notag
\end{equation}
From this bound, we may make several critical observations. Firstly, it is always possible to deterministically transform from one ensemble to another provided that the initial ensemble has $I(P;\Phi)$ greater than or equal to that of the target state. If, however, one wishes to transform to a state with larger $I(P;\Phi)$, one must do so probabilistically. Therefore, the cost of increasing $I(P;\Phi)$ is reduced probability of conversion between states. This codifies the resourcefulness of the probability-phase mutual information and provides an operational meaning to the quantity. 

\section{Conclusions}\label{sec:conclusions}

We have introduced the probability–phase mutual information $I(P;\Phi)$ and, employing a resource-theoretic approach, established it as a coherence measure for ensembles of pure quantum states. 
We have shown that $I(P;\Phi)$ quantifies a form of quantum coherence that exists only at the ensemble level: correlations across distributions of pure states rather than superpositions within individual states. The coherence surplus $\delta_{\mathcal{C}} = I(P;\Phi) + D_{\mathrm{KL}}(\mu_\Phi || \mathrm{unif}_\Phi) - \mathcal{C}(\rho) \geq 0$ formalizes the relationship between $I(P,\Phi)$, bounding how much ensemble structure survives the averaging into density matrices. The state conversion bound provides clear operational meaning to $I(P;\Phi)$, proving it is in fact a physical quantity.

Our framework is directed to the emerging trend of a ``physics of ensembles'' complementary to the standard density-matrix formalism, and addressing phenomena at the ensemble level: deep thermalization \cite{Mark2024MaximumEntropyDeep,Bhore2023DeepThermalizationConstrained} manifests as $I(P;\Phi) \to 0$ when projected ensembles approach the Haar distribution; geometric thermalization \cite{Anza2022GeometricQuantumThermodynamics} reveals temperature-tuned probability-phase correlations absent from thermal density matrices. More broadly, any scenario where the same mixed state arises from qualitatively different preparation procedures—adaptive measurements, engineered reservoirs, conditional dynamics—becomes accessible to ensemble-level analysis. As more experiments access the projected ensemble directly \cite{choiPreparingRandomStates2023, cotlerEmergentQuantumState2023}, tools that characterize its full distributional structure rather than only its low-order moments will become increasingly essential.

Ultimately, the probability-phase mutual information provides a powerful tool for an ensemble-first description of quantum coherence; one that complements density matrices by revealing the statistical architecture underlying mixed-state coherence.

\section*{Acknowledgments}

F.A. acknowledges several discussions on geometric quantum mechanics and information theory with J. P. Crutchfield. F.A. acknowledges financial support in the form of startup package from the University of Maryland, Baltimore County and from the Cybersecurity Institute at the University of Maryland, Baltimore County. F.A. and C.H. acknowledge useful feedbacks on this work by S. Deffner, A. Touil, and D. Girolami.

\bibliographystyle{quantum}

\bibliography{ProbabilityPhaseMutualInformation}

\onecolumngrid
\newpage
\appendix

\section{Proof of C1 - C6 properties}\label{appendix}

We provide detailed proofs that the probability-phase mutual information $I(P;\Phi)$ satisfies the six core properties required for a coherence measure in the ensemble framework.

\subsection{Proof of C1: Non-negativity}

\textbf{Property C1}: $I(P;\Phi) \geq 0$ with equality if and only if $\mu_{P,\Phi} \in \mathcal{I}_G$.

\begin{proof}
The mutual information is defined as the relative entropy:
\begin{align}
I(P;\Phi) = D_{KL}(\mu_{P,\Phi} || \mu_P \cdot \mu_\Phi)
\end{align}

Since relative entropy is non-negative, we have $I(P;\Phi) \geq 0$. Furthermore, $D_{KL}(\mu||\nu) = 0$ if and only if $\mu = \nu$ almost everywhere. Therefore, $I(P;\Phi) = 0$ if and only if $\mu_{P,\Phi} = \mu_P \cdot \mu_\Phi$, which is precisely the condition for $\mu_{P,\Phi} \in \mathcal{I}_G$.
\end{proof}

\subsection{Proof of C2: Monotonicity}

\textbf{Property C2}: For any free operation $\mathcal{K} \in \mathcal{F}_G$, we have $I(\mathcal{K}[\mu_{P,\Phi}]) \leq I(\mu_{P,\Phi})$.

\begin{proof}
Let $\mathcal{K} = \mathcal{K}_P \cdot \mathcal{K}_\Phi \in \mathcal{F}_G$ be a free operation. In the discrete case, these channels can be represented by conditional probabilities:
\begin{align}
K(p,\phi|\tilde{p},\tilde{\phi}) = K_P(p|\tilde{p})K_\Phi(\phi|\tilde{\phi})
\end{align}

For the continuous case, the action of $\mathcal{K}$ on a measure $\mu_{P,\Phi}$ with density $q(\tilde{p},\tilde{\phi})$ produces:
\begin{align}
\mathcal{K}[\mu_{P,\Phi}] \to \int d\tilde{p} \, d\tilde{\phi} \, K_P(p|\tilde{p})K_\Phi(\phi|\tilde{\phi})q(\tilde{p},\tilde{\phi})
\end{align}

Using the log-sum inequality:
\begin{align}
\sum_i a_i \log \frac{a_i}{b_i} \geq \left(\sum_i a_i\right) \log \frac{\sum_i a_i}{\sum_i b_i}
\end{align}

We can show that:
\begin{align}
D_{KL}(\mathcal{K}[\mu_{P,\Phi}] || \mathcal{K}_P[\mu_P] \cdot \mathcal{K}_\Phi[\mu_\Phi]) \leq D_{KL}(\mu_{P,\Phi} || \mu_P \cdot \mu_\Phi)
\end{align}

This follows from the data processing inequality for relative entropy.
\end{proof}

\subsection{Proof of C3: Strong Monotonicity}

\textbf{Property C3}: $I(P;\Phi)$ does not increase on average under selective free operations.

\begin{proof}
Consider a free instrument $\{\mathcal{G}_{i}\}$ such that each $\mathcal{G}_i$ is a sub-normalized free map and
$\mathcal{G}=\sum_i \mathcal{G}_i$ is a normalized free channel. Acting on a geometric state $\mu$,
\begin{align}
\mathcal{G}[\mu] = \sum_i \mathcal{G}_i[\mu].
\end{align}
Define the sub-normalized branch outputs
\begin{align}
\tilde{\mathcal{E}}_{i}^{PM} := \mathcal{G}_{i}[\mu],
\qquad
q_{i} := \tilde{\mathcal{E}}_{i}^{PM}(\Omega),
\end{align}
with $\Omega=\Sigma_{D-1}\times \mathbb{T}^{D-1}$, and for $q_i>0$ define the normalized post-measurement ensembles
\begin{align}
\mathcal{E}_{i}^{PM} := \frac{\tilde{\mathcal{E}}_{i}^{PM}}{q_{i}}.
\end{align}
Then
\begin{align}
\mathcal{G}[\mu] = \sum_{i}q_{i}\mathcal{E}_{i}^{PM}.
\end{align}

Now elevate the instrument by appending a classical register recording the outcome $i$. Define the flagged channel
$\mathcal{G}^{c}$ such that
\begin{align}
\mathcal{G}^{c}[\mu] = \sum_{i}q_{i}\mathcal{E}_{i}^{PM} \otimes \delta_{i} =: \mu^{c}.
\end{align}
Similarly define $\nu:=\mu_P\otimes\mu_\Phi$ and $\nu^c:=\mathcal{G}^c[\nu]$.

Using $I(\mu)=D_{KL}(\mu\|\nu)$ and contractivity of KL divergence under channels (data processing inequality),
\begin{align}
I(\mu)=D_{KL}(\mu\|\nu)\ \ge\ D_{KL}(\mathcal{G}^c[\mu]\ \|\ \mathcal{G}^c[\nu])
= D_{KL}(\mu^c\|\nu^c).
\end{align}

Write $\mu^c \sim \sum_i q_i f_i(p,\phi)\delta_i$ and $\nu^c\sim \sum_i r_i g_i(p,\phi)\delta_i$,
where $f_i$ is the density of $\mathcal{E}_i^{PM}$ and $g_i$ is the density of the normalized branch output
$\mathcal{G}_i[\nu]/r_i$ (which is a product density because $\nu$ is product and $\mathcal{G}_i$ factorizes).
Then
\begin{align}
D_{KL}(\mu^c\|\nu^c)
= \sum_i q_i \log\frac{q_i}{r_i} + \sum_i q_i D_{KL}(f_i\|g_i)
\ \ge\ \sum_i q_i D_{KL}(f_i\|g_i).
\end{align}
Since each $g_i$ is a product density, and mutual information satisfies
\begin{align}
I(\mathcal{E}_i^{PM}) = \min_{\alpha,\beta} D_{KL}(f_i\|\alpha\beta),
\end{align}
we have $D_{KL}(f_i\|g_i)\ge I(\mathcal{E}_i^{PM})$. Therefore,
\begin{align}
D_{KL}(\mu^c\|\nu^c) \ge \sum_i q_i I(\mathcal{E}_i^{PM}).
\end{align}
Combining with the contractivity step yields
\begin{align}
I(\mu) \ge \sum_i q_i I(\mathcal{E}_i^{PM}),
\end{align}
which proves strong monotonicity (C3).
\end{proof}

\subsection{Proof of C4: Convexity}

\textbf{Property C4}: $I(P;\Phi)$ is a convex function of the geometric quantum state.

\begin{proof}
From Cover and Thomas ~\cite{Cover2006ElementsInformationTheory}, relative entropy is jointly convex in its arguments. That is, for probability distributions $(p_1,q_1)$ and $(p_2,q_2)$:
\begin{align}
D_{KL}(\lambda p_1 + (1-\lambda)p_2 || \lambda q_1 + (1-\lambda)q_2) \leq \lambda D_{KL}(p_1||q_1) + (1-\lambda)D_{KL}(p_2||q_2)
\end{align}

Applying this to $I(P;\Phi) = D_{KL}(\mu_{P,\Phi}||\mu_P \cdot \mu_\Phi)$:

Let $\mu^{(1)}_{P,\Phi}$ and $\mu^{(2)}_{P,\Phi}$ be two geometric quantum states. Then:
\begin{align}
I(\lambda \mu^{(1)}_{P,\Phi} + (1-\lambda)\mu^{(2)}_{P,\Phi}) &= D_{KL}(\lambda \mu^{(1)}_{P,\Phi} + (1-\lambda)\mu^{(2)}_{P,\Phi} || \text{marginals}) \\
&\leq \lambda I(\mu^{(1)}_{P,\Phi}) + (1-\lambda)I(\mu^{(2)}_{P,\Phi})
\end{align}
\end{proof}

\subsection{Proof of C5: Uniqueness for Pure States}

\textbf{Property C5}: For pure states, $I(P;\Phi) = H(\Delta_G[\delta_{p,\phi}]) = 0$.

\begin{proof}
A pure state in the geometric framework corresponds to a Dirac measure:
\begin{align}
\mu_{P,\Phi} = \delta_{(p_0,\phi_0)} = \delta_{p_0} \cdot \delta_{\phi_0}
\end{align}

Since this already factorizes, we have:
\begin{align}
I(P;\Phi) = D_{KL}(\delta_{p_0} \cdot \delta_{\phi_0} || \delta_{p_0} \cdot \delta_{\phi_0}) = 0
\end{align}

The geometric dephasing of a pure state gives:
\begin{align}
\Delta_G[\delta_{(p_0,\phi_0)}] = \delta_{p_0} \cdot \delta_{\phi_0}
\end{align}

Both entropies are zero (for discrete measures with single support point), confirming the property.
\end{proof}

\subsection{Proof of C6: Additivity}

\textbf{Property C6}: For non-interacting systems, $I(P,P';\Phi,\Phi') = I(P;\Phi) + I(P';\Phi')$.

\begin{proof}
Consider two non-interacting systems with joint measure:
\begin{align}
\mu_{P,\Phi} \cdot \nu_{P',\Phi'}
\end{align}

The mutual information for the composite system is:
\begin{align}
&I(P,P';\Phi,\Phi') = D_{KL}(\mu_{P,\Phi} \cdot \nu_{P',\Phi'} || \mu_P \cdot \mu_\Phi \cdot \nu_{P'} \cdot \nu_{\Phi'})
\end{align}

Using the chain rule for relative entropy:
\begin{align}
D_{KL}(p \cdot q || r \cdot s) = D_{KL}(p||r) + D_{KL}(q||s)
\end{align}

We obtain:
\begin{align}
I(P,P';\Phi,\Phi') &= D_{KL}(\mu_{P,\Phi} || \mu_P \cdot \mu_\Phi) + D_{KL}(\nu_{P',\Phi'} || \nu_{P'} \cdot \nu_{\Phi'}) \\
&= I(P;\Phi) + I(P';\Phi')
\end{align}
\end{proof}

These proofs establish that the probability-phase mutual information $I(P;\Phi)$ satisfies all the required properties of a coherence measure in the ensemble framework, validating its use as a quantifier of quantum coherence for geometric quantum states.

\section{Proof of Coherence accounting Theorem }\label{apx:coherence-theorem}
We provide a proof of Theorem \ref{eq:coherence-surplus}.
\begin{proof}
Let the classical-quantum (cq) state $\Omega_{XQ}(\mu)$ have the classical label $X$:
\begin{equation}
    \Omega_{XQ}(\mu) := \int \mu_X(dx)\ket{x}\bra{x}\cdot \ket{\psi(x)}\bra{\psi(x)}
\end{equation}
where $\left\{ \ket{x}\right\}_x$ is an orthonormal basis for $X$. Tracing out the classical label produces the system density matrix:
\begin{equation}
    \rho = \mathrm{Tr}_X\Omega_{XQ}(\mu)
\end{equation}
Consider the pair of cq states $\Omega_{XQ}(\mu_{P\Phi})$ and $\Omega_{XQ}(\mu_P\cdot\nu_\Phi)$. They are block-diagonal in the classical register basis $\{\ket{x}\bra{x}\}$ and share the same conditional pure states $\ket{\psi(p,\phi)}\bra{\psi(p,\phi)}$. Thus the quantum relative entropy reduces to the geometric relative entropy:
\begin{equation}
    D_{KL}^Q(\Omega_{XQ}(\mu_{P\Phi})||\Omega_{XQ}(\mu_P\cdot\nu_\Phi)) = D_{KL}(\mu_{P\Phi}||\mu_{P}\cdot\mu_\Phi)
\end{equation}
Since the density matrix associated with the ensemble $\mu$ is computed via a partial trace (CPTP map), the quantum data-processing inequality can be invoked:
\begin{equation}
    D_{KL}^Q(\rho||\sigma) \le D_{KL}^Q(\Omega_{XQ}(\mu_{P\Phi})||\Omega_{XQ}(\mu_P\cdot\nu_\Phi)) = D_{KL}(\mu_{P\Phi} || \mu_{P}\cdot\nu_\Phi) \label{eq:quantum-data-processing}
\end{equation}
Now, letting $\nu_\Phi = \mathrm{unif}_\Phi$, and $\sigma = \mathrm{Tr}_X\Omega_{XQ}(\mu_P\mathrm{unif}_\Phi) = \Delta[\rho]$, the bound in Eq. \eqref{eq:quantum-data-processing} becomes:
\begin{equation}
    D_{KL}^Q(\rho||\Delta[\rho]) \le D_{KL}(\mu_{P\Phi}||\mu_P\mathrm{unif}_\Phi)
\end{equation}
Defining the coherence surplus as the difference in these relative entropies:
\begin{equation}
    \delta_{\mathcal{C}} = D_{KL}(\mu_{P\Phi}||\mu_P\mathrm{unif}_\Phi) - D_{KL}^Q(\rho||\Delta[\rho]) \ge 0
\end{equation}
Using the chain rule for classical relative entropy and identifying $\mathcal{C}(\rho) = D_{KL}^Q(\rho||\Delta[\rho])$:
\begin{equation}
    \delta_{\mathcal{C}} = I(P;\Phi) + D_{KL}(\mu_\Phi||\mathrm{unif}_\Phi) - \mathcal{C}(\rho) \ge 0
\end{equation}
\end{proof}

\section{Spiral Ensemble Benchmark}
To demonstrate non-zero mutual information, and benchmark our numerical algorithm, we introduce the spiral ensemble for a qubit: 
\begin{subequations}
\begin{align}
    p & \sim \text{Uniform}(0,1) \nonumber\\
    \phi & = 2\pi p + \varepsilon, \quad \varepsilon \sim \text{Uniform}(-\delta,\delta) \nonumber
\end{align}
\end{subequations}
This parametric ensemble allows us to explore correlations between probability and phase coordinates.
The noise parameter $\delta \in [0,\pi]$ controls the strength of statistical correlation between $P$ and $\Phi$, with $\delta=0$ leading to perfect linear correlations and $\delta \gg 1$ leading to complete independence caused by noise.

To compute the corresponding mutual information, we utilize the numerical scaling approach described above. We extract the information dimension $\mathfrak{D}$ and geometric entropy $H_G$ using the scaling of the joint and marginal entropies, as in Eq.\eqref{eq:entropy_scaling}

\begin{figure}[h!]
    \centering
    \includegraphics[scale=0.5]{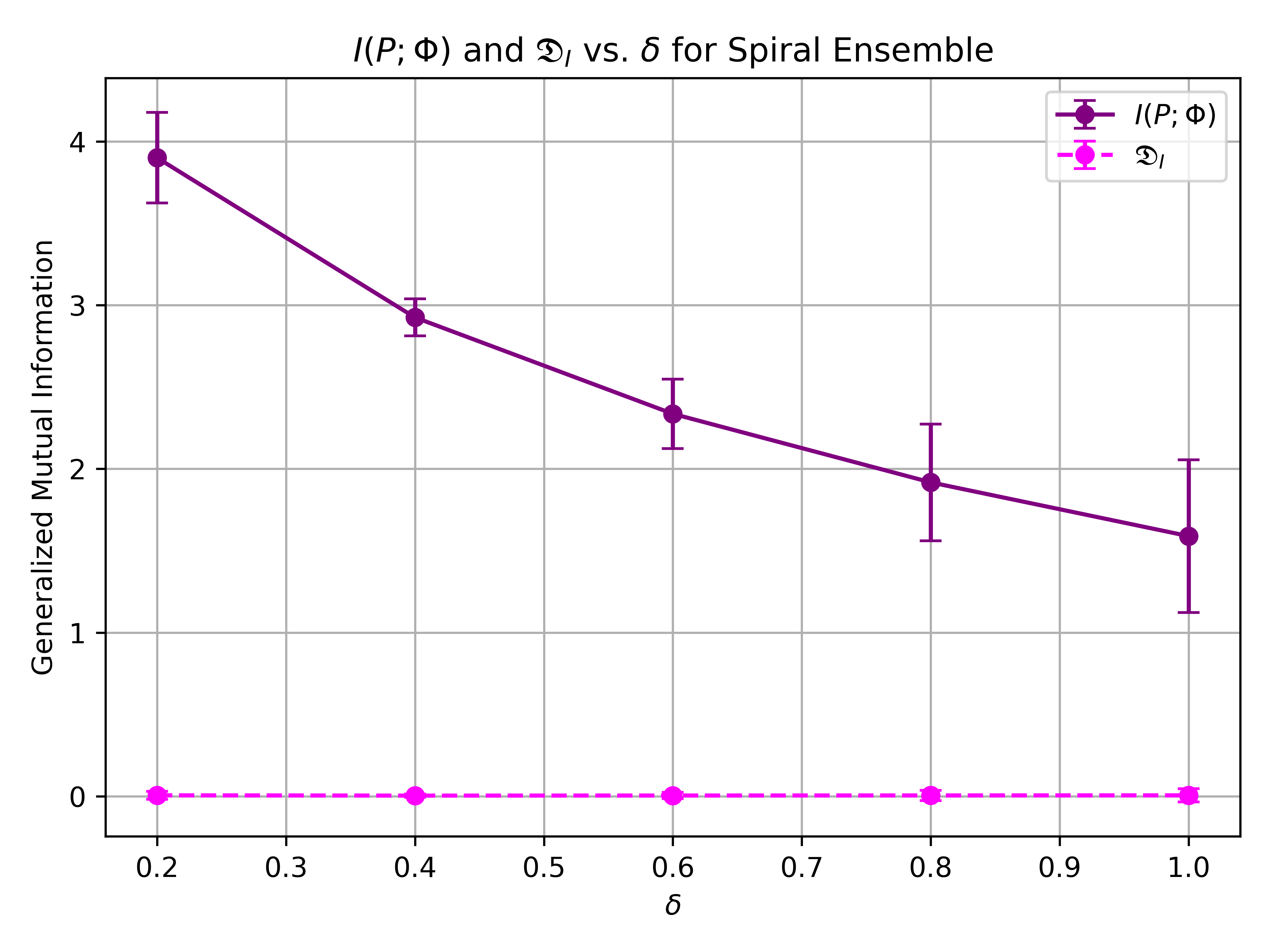}
    \caption{Probability–phase mutual information $I(P;\Phi)$ for the spiral ensemble as a function of noise strength $\delta$. For $\delta \ll 1$, the ensemble is near-perfectly correlated along a spiral curve, yielding finite mutual information. Adding noise gradually washes out these correlations, producing a monotonic decrease of $I(P;\Phi)$. As $\delta \to \pi$, the distribution approaches the uniform (Haar) ensemble and the mutual information vanishes. The information dimension $D_I \approx 0$ for all $\delta \neq 0$, indicating that the support dimension of the joint distribution matches that of the marginals.}
    \label{fig:spiral_ensemble_gmi_scaling}
\end{figure}

We then combine the results into the mutual information dimension $\mathfrak{D}_I$ and the geometric mutual information $I(P;\Phi)$ according to \eqref{eq:mutual_information_scaling}. We expect that for nonzero noise, $\mathfrak{D}_I \rightarrow 0$ since there is no deficit in support between the joint and the marginals. Also, as $\delta$ is increased, the spiral ensemble is broadened - causing it to approximate the Haar measure for sufficiently large $\delta$. From this, we can interpret the monotonic decrease in $I(P;\Phi)$ presented in Figure \ref{fig:spiral_ensemble_gmi_scaling} as an interpolation from the highly structured spiral ensemble at low noise to the Haar uniform measure, explaining the trend of $I(P;\Phi) \rightarrow 0$ for increasing $\delta$.

\end{document}